\def\spose#1{\hbox to 0pt{#1\hss}}
\def\approxlt{\mathrel{\spose{\lower 3pt\hbox{$\sim$}}
	\raise 2.0pt\hbox{$$<$$}}}
\def\approxgt{\mathrel{\spose{\lower 3pt\hbox{$\sim$}}
	\raise 2.0pt\hbox{$>$}}}

\def\multleft#1{\hbox to size{\vbox {\halign {\lft{##}\cr #1}}\hfill}\par}
\def\multright#1{\hbox to size{\vbox {\halign {\rt{##}\cr #1}}\hfill}\par}

\def\today{\ifcase\month\or January\or February\or March\or April\or May\or
      June\or July\or August\or September\or October\or November\or December\fi
      \space\number\day, \number\year}
\def\${\thinspace}

\def\boxit#1{\vbox{\hrule\hbox{\vrule\kern3pt\vbox{\kern3pt
          #1 \kern3pt}\kern3pt\vrule}\hrule}}

\def\Msun{\hbox{$\rm\thinspace M_{\odot}$}}

\documentstyle[psfig]{mn}
\begin{document}
\hsize=6truein

\renewcommand{\thefootnote}{\fnsymbol{footnote}}

\title{X-ray iron line variability for the model of an orbiting
flare above a black hole accretion disc}

\author[Mateusz Ruszkowski]
{\parbox[]{6.in} {Mateusz Ruszkowski}
\\
\footnotesize
Institute of Astronomy, Madingley Road, Cambridge CB3 0HA \\}

\maketitle
\begin{abstract}  
The broad X-ray iron line, detected in many active galactic nuclei, is
likely to be produced by fluorescence from the X-ray illuminated 
central parts of an accretion disc close to a supermassive black hole.
The time-averaged shape of the line can be 
explained most naturally by a combination
of special and general relativistic effects. Such line profiles contain
information about the black hole spin and the accretion disc
as well as the geometry of the emitting region and may help to test 
general relativity in the strong gravity regime.
In this paper we embark on the computation of the temporal response
of the line to the illuminating flux.
Previous studies concentrated on the calculation of
reverberation signatures from static sources illuminating the disc.
In this paper we focus on the more physically justified case of
flares located above the accretion disc and corotating with it. We compute
the time dependent iron line taking into account all general relativistic
effects and show that its shape is of very complex nature, 
and also present light curves accompanying the iron line variability.
We suggest that future X-ray satellites like XMM or
Constellation-X may be capable of detecting features present in the
computed reverberation maps.
\end{abstract}


\section{INTRODUCTION} 
\label{intro}
The most extreme continuum variability in AGN is usually seen in the X-ray
 spectral band.
The X-ray variability 
power spectrum reveals that low-amplitude fast flickering is an almost
permanent feature, but major changes, such as a flare with a large change
in luminosity (like doubling) occur much less frequently. The duration of such
flares may by of the order of a few thousand seconds (e.g. see figures in 
McHardy (1989)). Recent spectroscopic 
observations suggest that such X-ray flux variations can be accompanied by
Fe K$\alpha$ line variability which occurs on a similar time scale
(Iwasawa et al. (1996)). This might indicate that 
the size of the X-ray continuum emitting region is small in comparison
with other sources contributing to the overall continuum and that the iron 
line is produced in a compact region close to the continuum source
and is correlated with the X-ray outburst. The X-ray emission is 
thought to originate from the innermost part of the accretion disc. This region
 is subject to a number of instabilities which may produce flare-like activity.
There is a large body of observational and theoretical work which supports this
view. Among the proposed scenarios invoked to explain such a phenomenon are:
the amplification of magnetic fields in the disc by convection and 
differential 
rotation leading to the emergence of magnetic fields carrying hot plasma
(Galeev, Rosner \& Vaiana, 1979), 
 hot magnetic arcs due to the Parker instability 
(Chagelishvili, Leminadze \& Rogava (1989)), magnetic flares 
due to explosive release of stored magnetic energy in the accretion disc
 (Vries \& Kuijpers (1992)) and 
 vortices (Abramowicz et al. (1992)). The fluorescent 
iron line emission may be produced by irradiation of the disc material
by such flares located above the accretion disc (Lightman \& White 
(1988), George \& Fabian (1991), Matt, Perola \& Piro, (1991)).
As a result, the observed time-averaged spectra have distinctive, skewed, 
double-peaked profiles which reflect the Doppler and gravitational shifts
in a strongly curved spacetime (Fabian et al. 1989, Laor 1991,
Mushotzky et al. (1995), Tanaka et al. (1995), Nandra et al. (1997)). 
Apart from the line emission, the disc material
should also reflect a part of the X-ray continuum. Evidence for this effect 
was found by Pounds et al. (1990) for a number of Seyfert galaxies. 
Since the spectral properties of the Fe K$\alpha$ line are well known, the
iron line may be used as a probe of the very near environment of a black hole
and the black hole itself. This may enable one to 
study the geometry of the emission region, estimate black hole masses or
search for the dragging of inertial frames. The frame dragging effect is
especially interesting because its presence, in cooperation with the shear 
viscosity of an accreting material, may be responsible for 
the alignment of an accretion disc symmetry axis with the spin axis of a 
black hole or help to explain the stability of jets (Bardeen \& Peterson 
(1975)). \\
\indent
Much theoretical work has been done to 
search for the signatures of the mass and spin parameter of black holes
as well as the geometry of the line and continuum emitting regions.
Laor (1991) realized the importance of the impact of general relativistic
effects in the vicinity of the rotating black hole on the line shape.
This influence of the black hole spin on the iron line was later quantified
by Dabrowski et al. (1997). They calculated the iron line profile
assuming that the
line is produced in the thin corona above the accretion disc.
They considered the case of the Seyfert 1 galaxy MCG-6-30-15 and concluded 
that the very broad and skewed iron line can be explained if the 
emission originates from the disc extending down to the radius of marginal 
stability $r_{ms}<1.23$m which corresponds to the spin parameter $a>0.94$.
Thus they claimed that the black hole is rapidly rotating and should 
be described by Kerr geometry. 
A similar conclusion was drawn by Bromley, Chen, \& Miller (1997) who suggested
that the iron line profile was more likely to originate from a disc about a rotating
black hole.
Another approach was presented by Reynolds 
\& Begelman (1997) (hereafter RB97) who were able to explain 
the same data by assuming Schwarzschild geometry (i.e non-rotating black hole) 
and including the matter below the innermost stable orbit. However in their
model the iron line was excited by an external flare 
located on the symmetry axis above the accretion disc and illuminating
the regions below the innermost stable orbit. 
They also claimed that the rapidly rotating black hole in the radio-quiet 
active galaxy MCG-6-30-15 is contrary to some theoretical models invoked
to explain
radio loudness in AGN (e.g. Rees et al. (1982), Wilson \& Colbert (1995)).
However
more recent studies show that this is not necessarily the case 
(Ghosh \& Abramowicz (1997), Livio, Ogilvie \& Pringle (1999)).
Recently Young, Ross \& Fabian (1998)
calculated the spectrum corresponding to the RB97 model and found that it
should possess a photoelectric absorption edge of iron, which is not seen 
in the data. 
However with the quality of the current data the issue of the black hole 
spin remains ambiguous.
An entirely different approach to searching for the 
observational signatures of the black hole spin, mass and the geometry
of emission region was first suggested by Fabian et al. (1989) and
later considered in any detail by Stella (1990). He computed the iron line
variability from a disc around a Schwarzschild black hole assuming a pointlike
variable X-ray source located in the geometric centre of the disc.
For an observer at infinity any such variations of the primary X-ray source
would be 'echoed' by different locations on the disc leading to time evolution
of the line profile. The primary goal of this research was to suggest a method
to estimate black hole masses. His work was extended by 
Matt \& Perola (1992) and Campana \& Stella (1993, 1995) who considered
modified source geometries. Recently Reynolds et al. (1999) (hereafter RYBF99) 
generalized these 
results by assuming arbitrary black hole spin and searched for observational
signatures of this parameter. They mainly focused on the case of on-axis
source but briefly discussed preliminary results for a static 
flash-like off-axis flare in the vicinity of a Kerr black hole. \\
\indent
As noted above, 
the duration of bright flares can be of the order of $100m_{7}$ in geometric
time units ($1Gm_{7}10^{7}\Msun/c^{3}\approx m_{7}49 s$; 
for the black hole mass $M=10^{7}m_{7}\Msun$).
It is very likely that any variability associated with
blobs above accretion discs most likely comes from moving sources.
Thus in general, the assumptions of
a static primary X-ray source made in previous studies,
and in particular the assumption of an $\delta$-like flash from a static
flare made by RYBF99, limit the applicability of their results. 
In the present work we relax these assumptions and consider reverberation
effects from corotating flares above an accretion disc. We take into account
the effect of Doppler boosting of radiation from a moving flare which was
neglected by RYBF99 and allow the flares to revolve around
a black hole. 
We calculate the time dependent iron line for different values of 
inclination of the accretion disc and spin parameter (we consider the  
Schwarzschild and the almost maximal Kerr cases) and different
positions of the flare relative to the accretion disc.
We also compute the light curves accompanying the iron line variations
including the direct flux from the flare and the reflected flux 
from the disc. Similar calculations were performed for example by
Bao (1992), Karas et al. (1992),
Zakharov (1994) and Bromley et al. (1997) in the context of 
corotating spots located on the accretion disc. \\
\indent
The following Section and its subsections describe the assumptions made
in our model and the algorithm used to simulate the light curves and the 
iron line flux variations.
The last Section is devoted to the presentation and discussion of our results.
 
\section{The model assumptions and the algorithm}

\subsection{The X-ray flare and the disc illumination}
\label{flare}
We assume that the primary source of radiation is an X-ray flare situated 
above the accretion disc. As noted in the introduction, magnetic
 instabilities may cause such a flare 
to emerge from the disc and its position relative to the disc surface
may remain approximately unchanged for some time.
Thus we make the assumption that the flare is confined by the magnetic field
and corotates with the disc and
possesses fixed Boyer-Lindquist coordinates $r$ and $\theta$.
We also assume that: the flaring region is point-like, 
has a finite life-time and that photons propagate freely to either the 
observer, to the black hole, to the disc or escape to infinity
once emitted by the blob. This means that the corona
above the accretion disc is optically thin.
This is a generalization of the considerations
presented by RYBF99 who model X-ray flare as a
$\delta-$like impulse and additionally assume that radiation is produced
in the locally non-rotating frame of reference.
In order to numerically integrate the photon trajectory from the {\sl 
corotating} flare we first analytically derive appropriate constants of
 motion to propagate photons through the Kerr metric. All the necessary 
formulae are given in Appendix B, where we express the constants of 
motion in terms of polar and 
azimuthal angles in the local rest frame of the point-like source. 
Thus we can easily model the isotropic distribution of radiation 
in the emitter's
frame by a Monte Carlo method. This enables us to calculate the illuminating
flux as a function of time and energy 
in the rest frame of the corotating disc material taking into account
all the general relativistic effects, including previously neglected 
Doppler boosting from the moving flare. 
In the implementation of the algorithm we used the formula 
(see Appendix C):
\[
F(E_{d},\tilde{t}_{sd})\propto S_{E_{d}}=g_{sd}^{1+\alpha}E_{d}
^{-\alpha}\frac{f_{sd}}{\gamma d\phi dr}\left(\frac{\Delta}{A}\right)^{1/2},
\]
where we additionally assume that the flare emits a power law spectrum
with energy index $\alpha$. The factor $f_{sd}$, which is the ratio of
the number of photons intersecting a small patch on the disc 
(defined by $d\phi$ and $dr$) to the total
number of emitted photons, was calculated by means of the Monte Carlo method.
In the above formula $\tilde{t}_{sd}$ is the time it takes for 
photon emitted
from the source to reach the disc element and
$\gamma$ is the Lorentz factor for the relative motion of the disc element
and the locally non-rotating observer. $g_{sd}$ is the redshift factor given by
\[
g_{sd}=\frac{u_{d}^{\mu}p_{d\mu}}{u_{s}^{\mu}p_{s\mu}} ,
\]
where $p_{d\mu}$ and $p_{s\mu}$ are the photon four-momenta at the disc and
source respectively. We follow the lines of reasoning of Cunningham (1975) and
split the velocity field of the accretion disc $u_{d}$
into the region outside the radius of marginal stability,
where matter follows circular orbits, and the region within the innermost 
stable orbit where the matter has a negative radial component and 
spirals towards the black hole. The formulae for the flare velocity
field $u^{\mu}_{s}$ are collected in Appendix B.

\subsection{Reprocessing of radiation by the accretion disc}
\label{disc}
The radiation which impinges upon the disc surface is reprocessed and 
reflected. The shape of the reflected spectrum depends upon many factors, 
the most important being the disc structure and ionization state of the
accretion disc. In the subsequent discussion we assume the disc to be 
optically thick and geometrically thin.
The optical depth of the accretion disc was calculated 
by many authors (e.g. Frank, King, \& Raine. (1995), RB97)
and is shown to be large. This guarantees that the 
illuminating radiation can be efficiently reflected. However the albedo
of the reflecting material depends on the ionization parameter
which varies across the disc surface: 
$\xi(r,\phi)=4\pi F_{X}(r,\phi)/[n(r)\cos n]$, 
where $n(r)$ is the comoving electron number density given by Eq. 4 of 
RYBF99, $\cos n$ is the angle between the incident rays and the normal
to the disc surface in the comoving frame of the disc 
and $F_{X}$ is the X-ray flux defined over some fixed energy band
at a particular position on the disc. We calculate 
the parameter $\xi$
and use the results of Ross, Fabian \& Young (1998)
to calculate the albedo $a_{R}(E)$ as a function of energy
for all grid elements on the disc. 
We then adopt the simple prescription of RYBF99 to separately
model the iron line fluorescence in the rest frame of the moving 
disc element, i.e. we assume that for $\xi<100$ erg cm s$^{-1}$ the matter
emits cold iron line at 6.4 keV with some yield $Y$, for $\xi$ in the range
from 100 erg cm s$^{-1}$ to 500 erg cm s$^{-1}$ there is no emission because 
the radiation is destroyed by Auger mechanism, for $\xi$ between 
500 erg cm s$^{-1}$ and 5000 erg cm s$^{-1}$ we use a composition of two
lines at 6.67 keV and 6.97 keV each with the fluorescent yield $Y$ and 
assume that no iron line is produced for $\xi>5000$ erg cm s$^{-1}$ because
of total ionization of the matter (here we approximate the iron lines as delta 
functions). Note that these additional complications
are essentially relevant only for the Schwardschild 
black hole because only in this case can a significant fraction of the disc
material within the innermost stable orbit ($r_{ms}=6m$)
be in the higher ionization state (large $\xi$).
This arises from the fact that the radial density gradient of the 
matter spiralling towards the black hole is strongly negative in this region
so the disc significantly decreases its density whereas
the density changes for radii greater than $r_{ms}$ are comparatively 
negligible and the disc remains relatively dense and thin, 
at least for small accretion rates. 
This means that the disc may be 
treated as approximately cold for $r>r_{ms}$ and the variations of the 
ionization parameter induced by the density changes should not dramatically
alter the reflection from the disc in this region.
Thus the assumption of RYBF99 about the high density
and negligible ionization of the region outside $r_{ms}$ in the Schwarzschild
case may be 
justified. In the Kerr case the radius of marginal 
stability asymptotically approaches the event horizon and we assume that 
the disc may be 
approximated as being cold on average. The results of R\'o\.za\'nska 
et al. (1998),
who show that the presence of a hot corona makes the disc denser and 
less ionized, further strengthen the above argument.\\
\indent
In the subsequent considerations we use the additional simplifying assumption
of isotropic reflection and write the reflected intensity in the form:
\[I_{R}(E_{d},\tilde{t}_{sd})=\frac{1}{\pi}a_{R}(\xi,E_{d})F(E_{d})\]
\subsection{Calculation of light curves and iron line variability}
In order to calculate the time-varying iron line profiles and 
the corresponding continuum light curves we used the following prescription:\\
\indent
1) We use a ray-tracing technique to follow the trajectories of photons 
from positions $(x,y)$ on the observer's image plane to the disc,
keeping track of all Boyer-Lindquist coordinates $\tilde{t}_{do}(x,y)$,
 $r(x,y)$, $\theta(x,y)$, 
 $\phi(x,y)$, until the photons either intersect the accretion disc plane
or disappear below the event horizon (The observer is located at 
$r=1000m$, $\theta=i$ and $\phi=0^{o}$, where $i$ is the inclination).
 We then calculate the redshift 
factor corresponding to the particular final position of a photon and 
its arrival time from the disc. We generate an image of approximately
$1600\times 1600$ pixels covering roughly an area of $160m\times 160m$.\\
\indent
2) Using the time dependent ionization and illumination pattern 
described above and albedo $a_{R}(E,\xi)$, 
we specify the emergent flux from the disc 
as a function of time and energy in the fixed spectral band,
i.e. we integrate all photons coming from regions of equal arrival time   
regions on the accretion disc at a particular energy.
In doing so we also take into account the fact that the source changes 
its position and rotate the illumination and ionization patterns
accordingly. This allowed us to compute the reflected spectrum as a function 
of time and thus the reflected component of the light curve. We use the
following formulae (see appendix C for additional explanations):
\[F_{R}(t)=\sum_{n} F^{(n)}_{R}(t),\]
\noindent
where $F^{(n)}_{R}(t)$ is the contribution to the flux from the flare at the
n-th position given by:
\[F^{(n)}_{R}(\tilde{t})=\frac{1}{\Delta \tilde{t}}
\sum_{\tilde{t}\rightarrow \tilde{t}+\Delta \tilde{t}}C(E_{o1},E_{o2},\alpha)
\int \int_{E_{o1}}^{E_{o2}}g_{do}^{3}I_{R}(E_{o}/g_{do})
dE_{o}\frac{dxdy}{r_{o}^{2}},\]
where $g_{do}$ is the redshift factor for the disc-to-observer case,
$dxdy$ is the pixel surface area on the observer's image plane, 
$r_{0}$ is the distance from the black hole system and $\tilde{t}=
\tilde{t}_{sd}+\tilde{t}_{do}$.
\\
\indent
3) We compute the direct flux component by propagating
photons from the source to the observer. We use a formula analogous 
to that above to calculate this flux.\\
\indent
4) Having specified the illumination pattern and ionization state 
for a given position of the flare above the accretion disc,
we determine the contribution of each grid element on the disc surface
to the iron line profile from the formula:
\[
\Psi^{(n)}(E,\tilde{t})\propto \frac{1}{\Delta \tilde{t} \Delta E}
\sum_{\tilde{t}\rightarrow \tilde{t}+
\Delta \tilde{t}, E\rightarrow E+\Delta E} 
Y(\xi,E/g_{do})g_{do}^{4}F_{X}(r,\phi),
\]
where $F_{X}(r,\phi)$ is the integrated over a fixed energy band X-ray 
flux at the given position on the disc.
The above function is the Green's function (transfer function) 
corresponding to the n-th position of the flare.\\
\indent 
5) In the last step we use the transfer functions for single isolated flares
in order to calculate the observed iron line flux variations. We use the 
following expression:
\[F_{Fe}(E,t)\propto\sum_{n} \Psi^{(n)}(E,t),\]
where we also take into account the delays due to the revolution of the flare
around the black hole.
As in the case of the reflected flux, we 
interpolate the contributions to the flux 
from the isolated flares and then add the interpolated fluxes.
Note that the binning of the transfer
functions in time would introduce a small additional (artificial) flux
variability, because we physically rotate the illumination pattern.
This is because the differences between the arrival times from
the neighbouring disc elements on the approaching side of the disc would
be shorter compared to the corresponding differences on the opposite side.

\section{results}
We present iron line flux variations
and the corresponding light curves for one full revolution of the flare 
around the black hole with a step-like luminosity variation in the 
local frame of reference of the flare. This is for illustrative
purposes and we stress that while it is trivial, for example, to generalize
the code to include arbitrary intrinsic luminosity variations in the rest 
frame of the flare, we do not wish to complicate the
interpretation of the results. Future
observations may guide us as to how to 
introduce additional modifications in the computations. The results
in the present form can in principle be applied to the case of isolated
flares or at least strong outbursts accompanied by minor ones.  
In the computations we assume that the source emits a power law spectrum
with energy index $\alpha=1$.

\begin{figure}
\centerline{\psfig{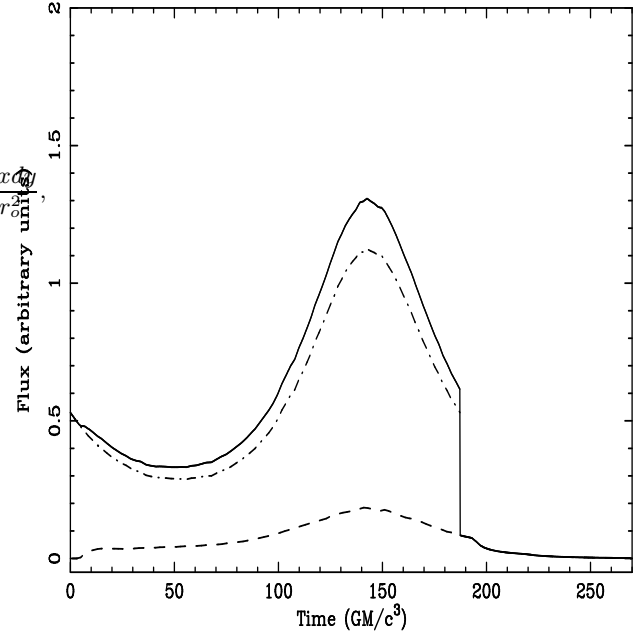}}
\caption{Total light curve (solid line), the direct flux from the flare (dot-dashed line) and the small 
contribution from the reflected component (dashed line)
for the same set of parameters as in Fig. 3. The flare is switched on from 
$\phi=0^{o}$ to $\phi=360^{o}$.}
\end{figure}

\begin{figure}
\centerline{\psfig{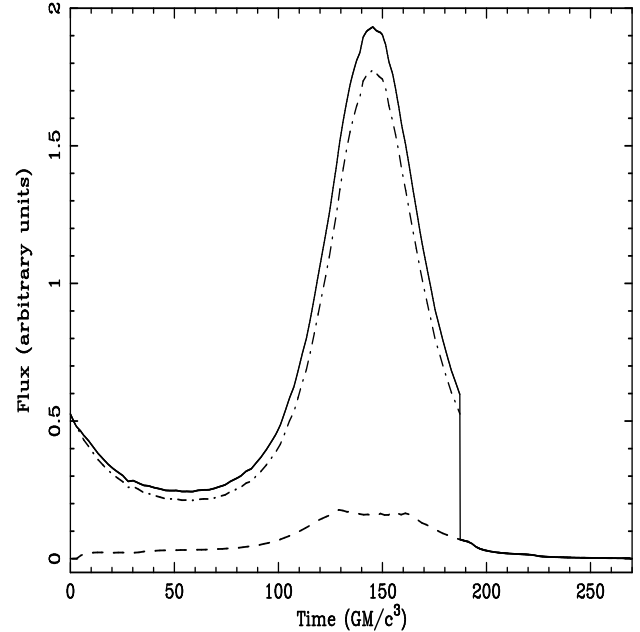}}
\caption{Total light curve (solid line), the direct flux from the flare (dot-dashed line) and the small 
contribution from the reflected component (dashed line)
for the same set of parameters as in Fig. 4. The flare is switched on from 
$\phi=0^{o}$ to $\phi=360^{o}$.}
\end{figure}

Fig.1 and Fig.2 present the light curves.
 The light curves are a superposition of the direct
flux from the flare and the reflected flux from the disc. The small offset 
between these two components (note the shift between initial and final 
times of the reflected and direct flux) is due to the additional
time photons need to propagate towards the disc and back. The small albedo
of the disc material in the chosen energy band (2-9 keV) results in 
a relatively small contribution from reflection continuum 
to the total light curve.
However the albedo strongly increases with energy and thus the reflection 
would be much more efficient in the harder part of the observed spectrum.
Another effect which would enhance the reflected component (and suppress
the direct one) is the focusing of light from the flare. More radiation 
from the flares closer to the central black hole would be focused towards
the disc and reflected, and simultaneously less direct 
flux would reach the observer
(Martocchia \& Matt, (1996)). The peaks in the direct light curve are mainly
attributed to the Doppler effect. Note how the flux variability is enhanced
by the inclination of the system. 
For higher inclinations the Doppler effects are more pronounced and the peaks
are narrower and have higher maxima. It is important to note that
the flux variability observed in the chosen spectral band (2-9 keV in our case)
is also a function of the energy index and
will be increased in the case of the steeper spectra. 
The peaks due to gravitational focusing
(e.g. Cunningham \& Bardeen, (1973)) are not seen here because we limit our
discussion to low inclinations in order to avoid additional
complications related to the limb darkening. This is not a serious limitation 
to our model because the majority of Seyfert 1 galaxies are likely 
to have small inclination angles. However 
we compare our results for the direct 
flux in the test case of a source orbiting the black hole in the equatorial
plane for very high inclinations of the orbit and find agreement with the
results of Cunningham \& Bardeen (1973). \\
\indent
\begin{figure}
\centerline{\psfig{figure=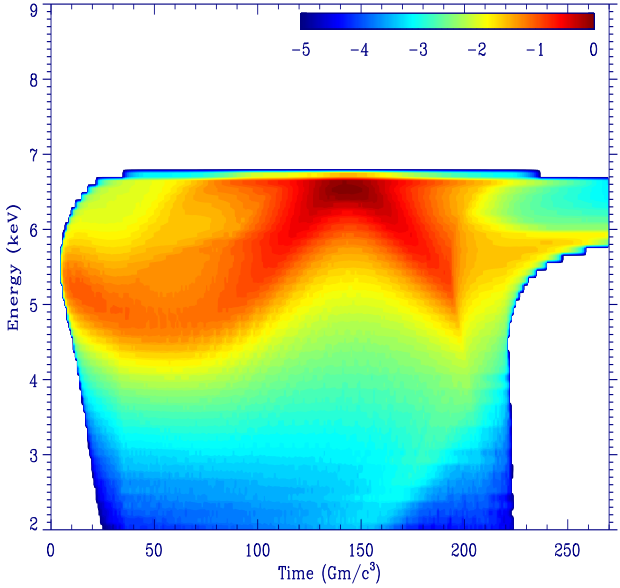,width=0.5\textwidth,height=0.35\textheight,angle=0}}
\caption{The colour map represents time-sequence of the iron line spectrum, i.e. 
$\log (F($time,energy$))$, where $F$ is the flux divided by the maximum flux during the whole
time interval. The Boyer-Lindquist coordinates of the flare are:
$\theta_{flare}=70^{o}$, $r_{flare}=10m$ with  $\phi_{flare}$ changing from $0^{o}$ to $360^{o}$.
The spin parameter and the inclination are $a=0.998$ and $i=30^{o}$ respectively.}
\end{figure}
\begin{figure}
\centerline{\psfig{figure=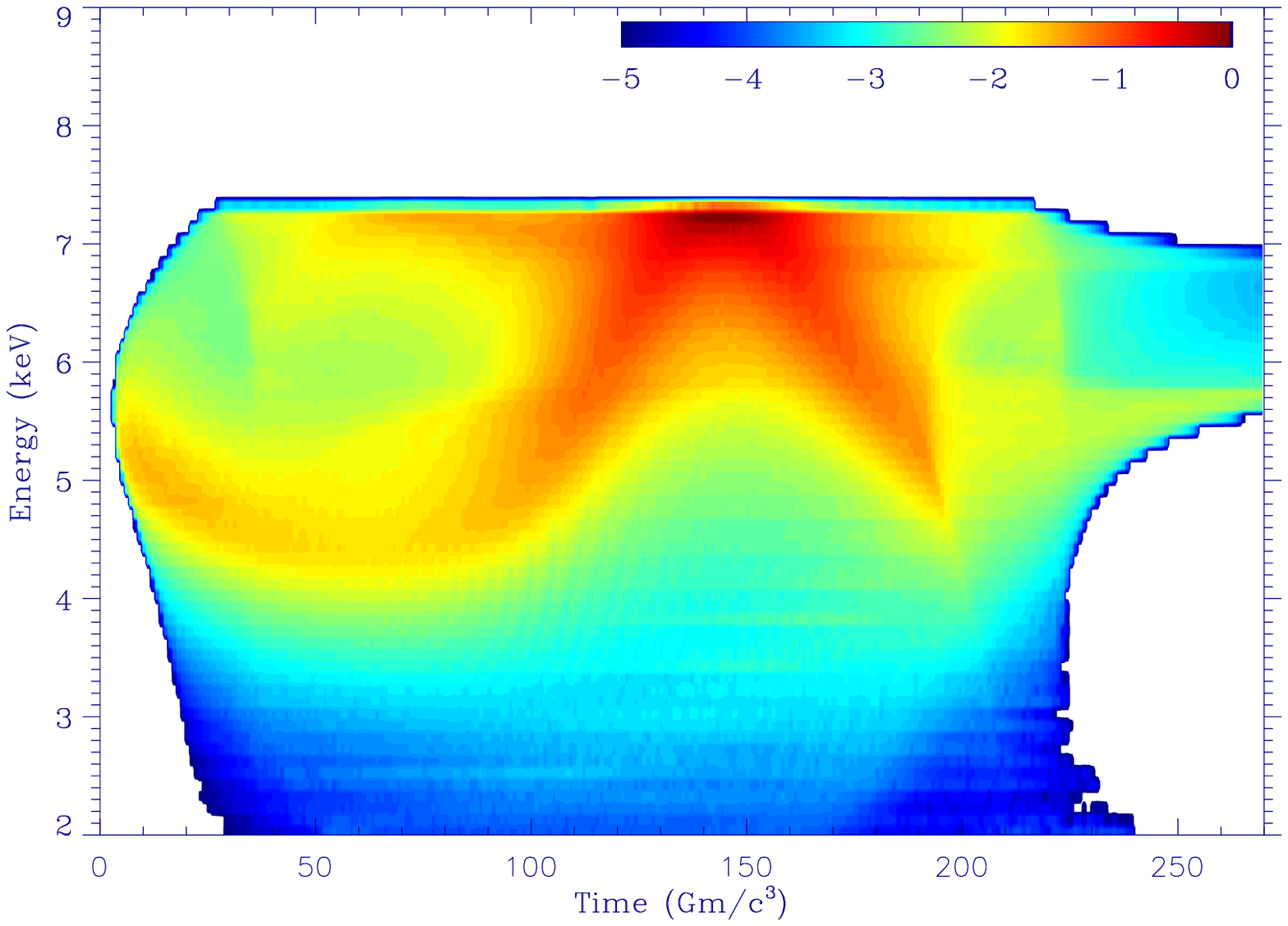,width=0.5\textwidth,height=0.35\textheight,angle=0}}
\caption{As Fig. 3, but for $i=50^{o}$.}
\end{figure}
\begin{figure}
\centerline{\psfig{figure=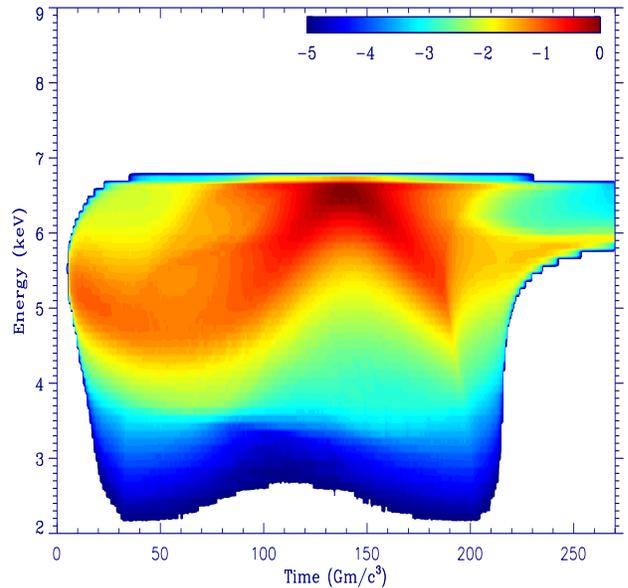,width=0.5\textwidth,height=0.35\textheight,angle=0}}
\caption{As Fig. 3, but for $a=0.0$.}
\end{figure}
\begin{figure}
\centerline{\psfig{figure=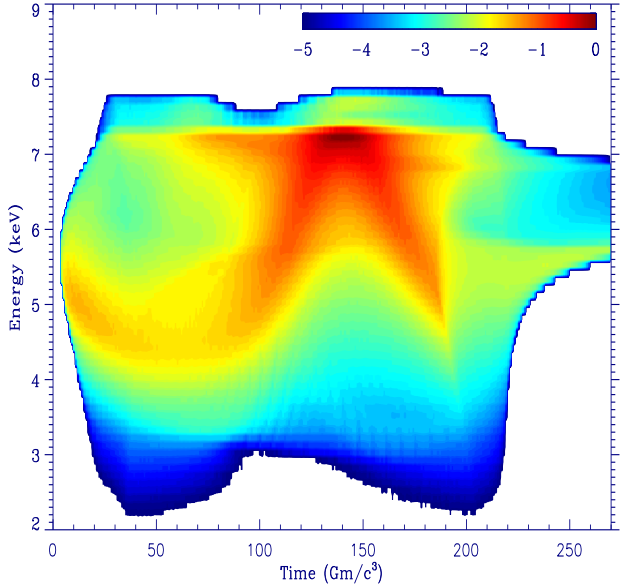,width=0.5\textwidth,height=0.35\textheight,angle=0}}
\caption{As Fig. 3, but for $a=0.0$ and $i=50^{o}$.}
\end{figure}
\begin{figure}
\centerline{\psfig{figure=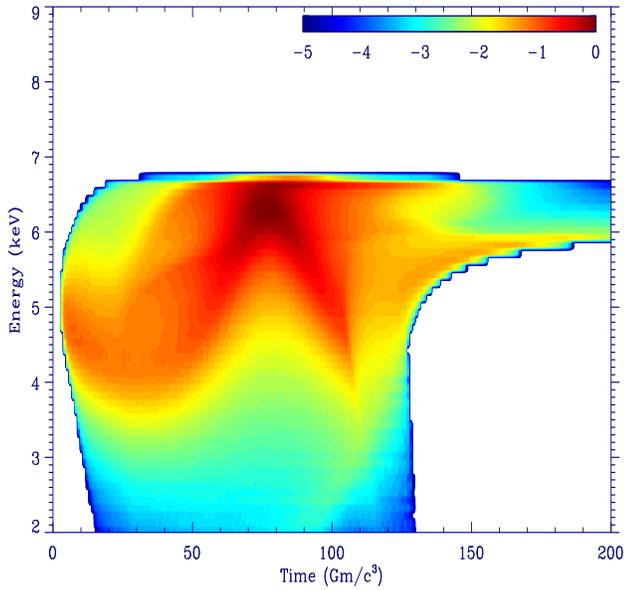,width=0.5\textwidth,height=0.35\textheight,angle=0}}
\caption{As Fig. 3, but for $r_{flare}=6.5m$.}
\end{figure}
\begin{figure}
\centerline{\psfig{figure=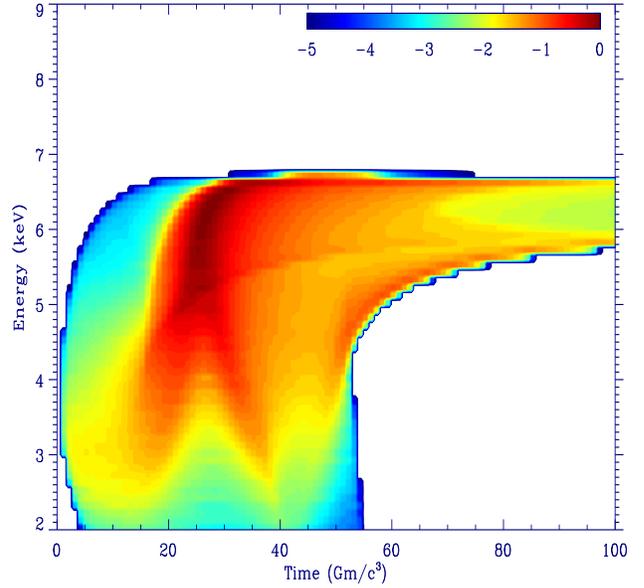,width=0.5\textwidth,height=0.35\textheight,angle=0}}
\caption{As Fig. 3, but for $r_{flare}=3m$.}
\end{figure}
\begin{figure}
\centerline{\psfig{figure=30_k_10.ps,width=0.5\textwidth,height=0.35\textheight,angle=270}}
\caption{Integrated iron line profile for $t=(70\div 80)Gm/c^{3}$
 after the detection of the first
signal from the flare for: $a=0.998$, $\theta_{flare}=70^{o}$, $r_{flare}=10m$
and $i=30^{o}$}
\end{figure}
\begin{figure}
\centerline{\psfig{figure=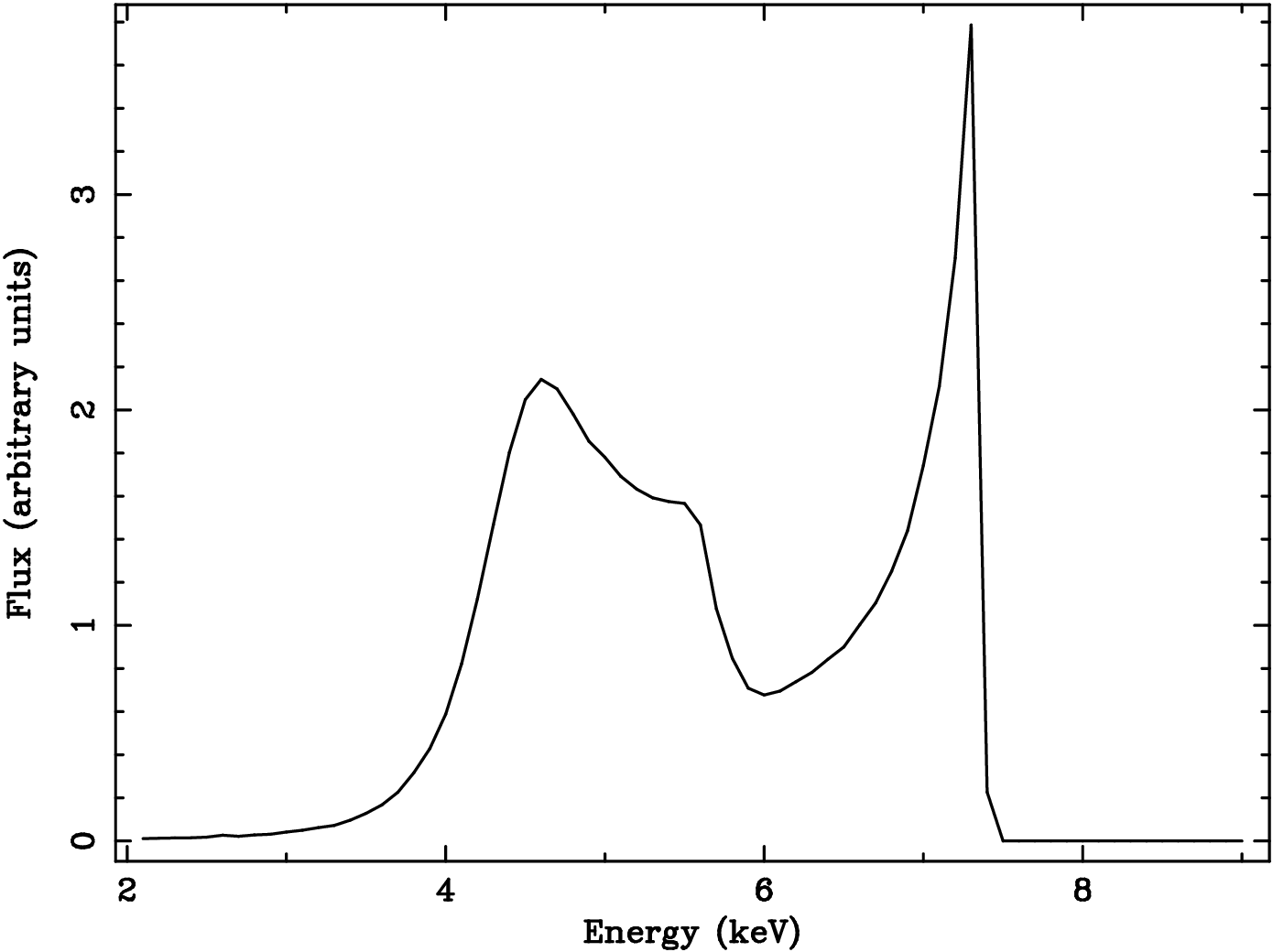,width=0.5\textwidth,height=0.35\textheight,angle=270}}
\caption{Integrated iron line profile for $t=(70\div 80)Gm/c^{3}$
 after the detection of the first
signal from the flare for: $a=0.998$, $\theta_{flare}=70^{o}$, $r_{flare}=10m$
and $i=50^{o}$}
\end{figure}
\begin{figure}
\centerline{\psfig{figure=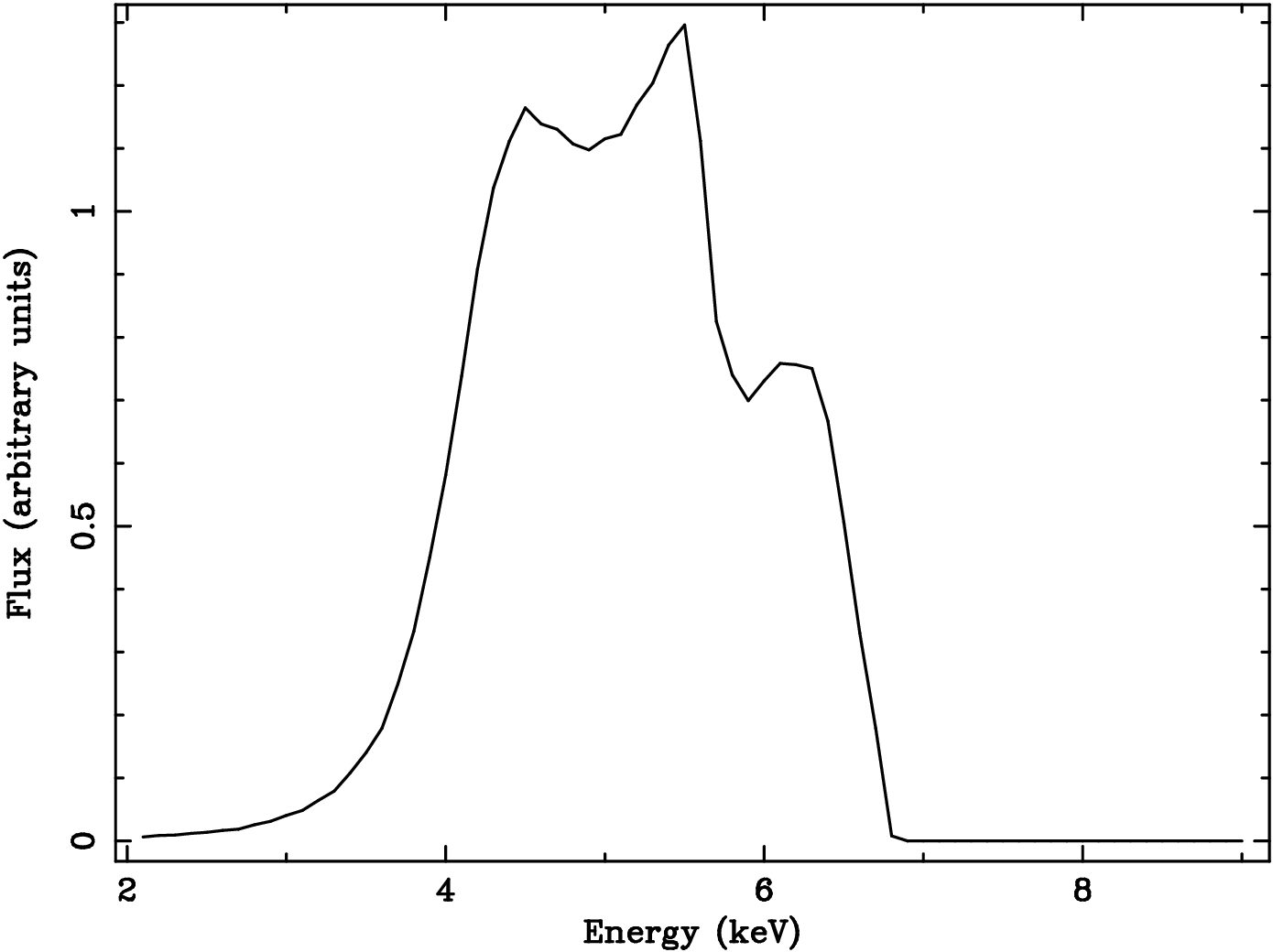,width=0.5\textwidth,height=0.35\textheight,angle=270}}
\caption{Integrated iron line profile at $t=(40\div 45)Gm/c^{3}$
 after the detection of the first
signal from the flare for: $a=0.998$, $\theta_{flare}=70^{o}$, $r_{flare}=6.5m$
and $i=30^{o}$}
\end{figure}
\begin{figure}
\centerline{\psfig{figure=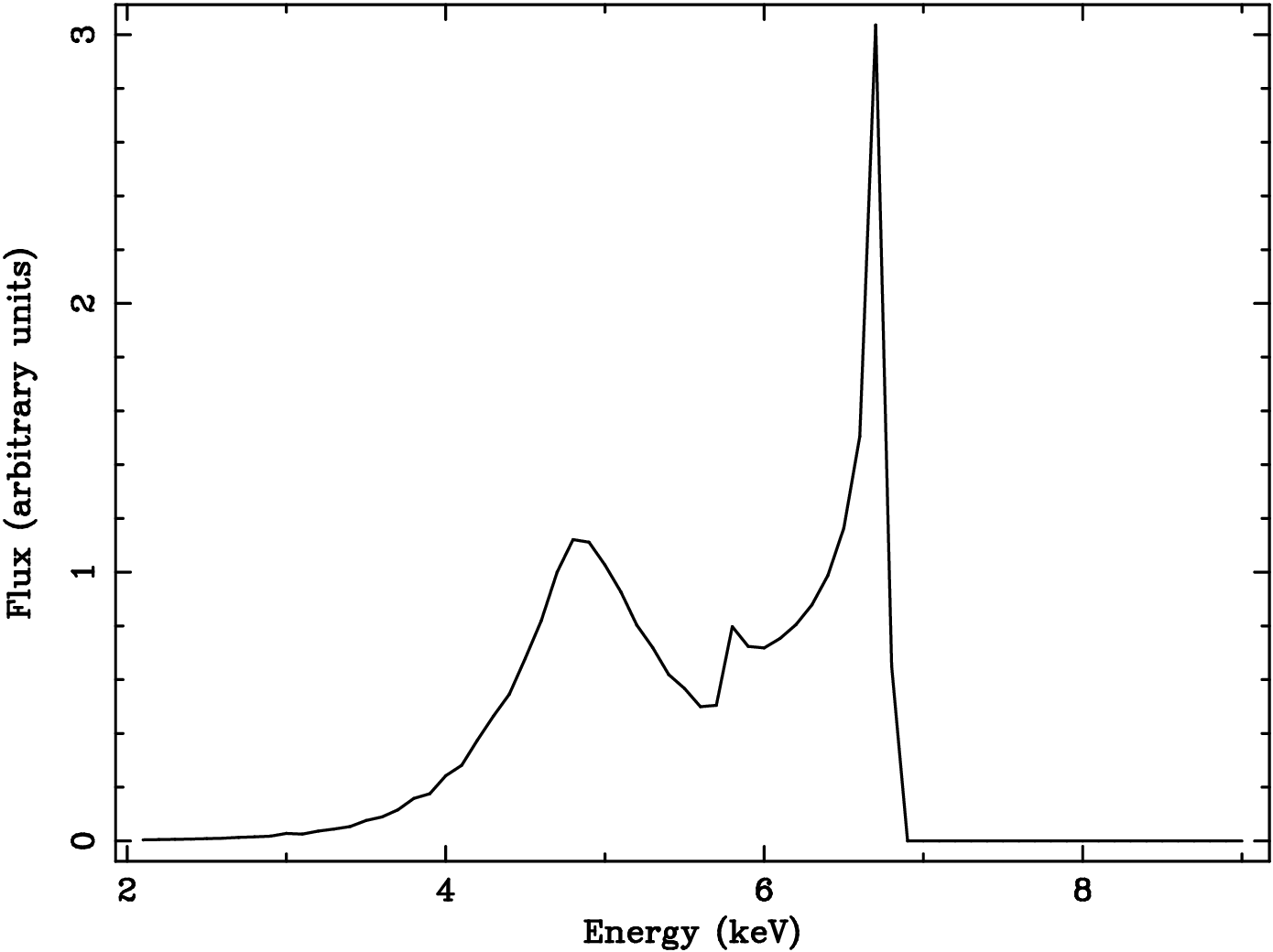,width=0.5\textwidth,height=0.35\textheight,angle=270}}
\caption{Integrated iron line profile at $t=(50\div 55)Gm/c^{3}$
 after the detection of the first
signal from the flare for: $a=0.998$, $\theta_{flare}=70^{o}$, $r_{flare}=3m$
and $i=30^{o}$}
\end{figure}
Figs from 3 to 8 show the evolution of the iron line spectra after
the subtraction of the varying power law continuum for different sets of
parameters. 
As for the case of the light curves, the time offset of the initial
spectrum is the difference between the arrival of the direct continuum
flux and the first appearance of the iron line. There is a clear 
difference between the results for the two considered extremal values
of the spin parameter (e.g. compare Fig. 4 and Fig.6).
 The iron line extends to lower frequencies in the 
Kerr case which is due to the cold iron line being produced very close to
the black hole. The influence of the ionization of the accretion disc 
below the radius of marginal stability $r_{ms}=1.23m$ is 
relatively unimportant. On the contrary, in the case of the Schwarzschild
black hole the bulk of the cold iron line is produced only above 
$r_{ms}=6m$ and accretion disc may be in highly ionized within
this radius. Thus for $r < 6m$ the line may be either destroyed by the 
Auger process or not produced because of the total ionization of the disc 
material. The radiation from this region would be subject to high energy
shifts which means that the lack of the iron line flux at lower energies 
(see Fig. 5 and Fig.6) is the result of the fact that some parts of
the disc below $r_{ms}=6m$ cannot produce the iron line.
For the flare at the position $\phi_{flare}$ 
the regions on the disc responsible for the production of the 
iron line in the higher ionization states (500 erg cm s$^{-1}$ $< \xi <$ 5000 
erg cm s$^{-1}$) are located around $\phi=\phi_{flare}\pm 90^{o}$. The regions
on the opposite side of the black hole ($\phi=\phi_{flare} +180^{o}$)
and nearest
to the flare ($\phi =\phi_{flare}$) have even 
higher ionization parameter which is the 
result of the strong light bending and the direct illumination of the disc
surface respectively. This effect gives rise to the varying width of the line
for $a=0$ with a minimum around $(100\div150) Gm/c^{3}$ 
(see Fig.5 and Fig. 6). 
Future high-throughput instruments like XMM and Constellation-X may be capable of detecting such feature in the temporal iron line profiles.
However we stress that a detection of this weak feature would be difficult
because it could be overwhelmed by the continuum variations.
Other interesting features
are the two 'bumps' seen for the inclination $i=50^{o}$ and the Schwarzschild
black hole. Here the iron line extends to higher frequencies
than the maximum frequency in the Kerr case 
for the same inclination of the disc (compare Fig. 4 and Fig.6).
 These features are due to 
the highly blueshifted 'hot' iron lines ($E^{(1)}_{rest}=6.67$ keV and
$E^{(2)}_{rest}=6.97$ keV)
from the ionized regions. This effect should also
be within the reach of capabilities of XMM and particularly Constellation-X.
It has to be stressed however that
the above results are sensitive to the 
parameters and depend on the position of the flare and its X-ray efficiency 
(here assumed to be $\eta_{X}=0.06$)
and the density of the disc below the innermost stable orbit
(which is regulated by the half thickness of the disc $h_{disc}$; 
here we assumed $h_{disc}=0.05 r$). In principle, a source of very
low efficiency located at a very large distance and large $\theta_{flare}$
and illuminating the high density region below the innermost stable orbit
could not ionize it, and a faint cold iron line could also be
produced all the way
down to the event horizon in the Schwarzschild case.
In such situation it would be difficult to 
distinguish between the Kerr and Schwarzschild black holes. 
On the other hand, the flares located closer to the black hole
and closer and less inclined
with respect to the rotation axis would ionize the regions below the innermost 
orbit much more efficiently leading to even more pronounced differences
(e.g. then those shown in Fig.4 and Fig.6)
between the temporal iron line profiles for the two extreme values
of the spin parameter.
The work of RB97 and Young et al. (1998) has shown that in the special case of 
a nonrotating black hole and the flare located on the rotation axis
very close to the black hole (e.g. $h=3.5m$) essentially all radiation
is focused towards the disc (and the black hole) leading to strong ionization.
Young et al. (1998) also demonstrated that in this case the iron absorption
edge becomes important. This effect leads to the trough in the iron line 
spectrum above the considered iron line rest energies.
Thus more accurate modelling should include the absorption edge which
would decrease the iron line flux from the ionized regions below $r_{ms}$
and create an oscillating absorption feature in the reverberation maps. 
It is also possible that other processes could influence the iron line
production below $r_{ms}$.
For example, the radiation drag could decrease the momentum of the infalling 
matter and thus accelerate the accretion below $r_{ms}$ leading to a decrease
in density and an increase in the ionization parameter in this region. 
Thus, as a result of the above processes, the iron line 
in the Schwarzschild case may not extend to  
frequencies as low as those in the Kerr case and may possess the absorption
features, so in general the appearance of the 
reverberation maps for the two cases should indeed be different, 
making it possible to distinguish between a rotating and nonrotating 
black hole.\\
\indent
The most prominent feature seen in all diagrams showing the evolution
of the iron line spectrum is the drifting maximum of 
the flux. This sinusoid-shaped feature is due to the fluorescence
from the most 
strongly illuminated part of the disc located just below the flare. 
Its shape and brightness is the result of
special and general relativistic effects. Initially the flare moves in a
direction perpendicular to the line of sight and the maximum is redshifted
due to the transverse Doppler effect and gravitational frequency shift.
Later it gradually begins to 
recede from the observer. The brightness of this maximum 
is suppressed at this stage because
the flux from the highly illuminated part of the disc below the flare
decreases as a result of light beaming in the 
opposite direction relative to the observer. 
The other interesting
feature seen in this phase is the double maximum clearly seen 
around $(50\div100) Gm/c^{3}$ in Fig.4 and Fig.6.
As mentioned above, the lower maximum is the result of the 
redshifting of the bulk of the radiation from below the receding flare.
The emerging upper maximum on the other hand, is due to fluorescence from
the illuminated regions on the approaching side of the disc.
As the flare rotates and enters the approaching side of the disc the 
bulk of the iron line flux shifts towards high energies and the 
Doppler boosting of the radiation reflected by the strongly illuminated 
regions of the disc leads to the rise of the iron line maximum and its 
brightening. After the flare passes through $\phi_{flare}\approx 270^{0}$
the energy shift decreases and gravitational and transverse Doppler shifts
gradually take over. Note that during the whole revolution of the flare 
the redshift factor $g<1$ for a longer time, because the gravitational 
and transverse Doppler shifts may for some time overwhelm the blueshifts
when the flare is on the approaching side of the disc, i.e. when the flare
has just entered or is just about to leave the approaching side of the disc.
This effect gives rise to a flattening of the minimum (in energy) 
of the main flux maximum, i.e. the bright $\Lambda$-shaped feature appears
to be sharper (e.g. see Fig. 4 and Fig. 6).\\
\indent
A very interesting feature can be seen when the flare is receding 
($\phi\approx 90^{o}$). The main redshifted flux maximum broadens 
(see Fig. 4 and Fig. 6) or even splits into two maxima (see Fig. 3, Fig. 5
or Fig. 7). In order to illustrate this effect better, appropriate 
cross-sections through the reverberation maps shown in Fig. 3 and Fig. 4  
are given (see Fig. 9 and Fig. 10 respectively). Note that in the case 
of Fig. 9 three maxima are seen.
This thickening or splitting is a purely general
relativistic effect. The middle maximum is a consequence of strong bending
of light which is focused on the opposite side of the black hole relative
to the actual position of the flare and leads
to the enhancement of the illumination of the disc in this region.
The influence of this effect is even more important for the flare 
located closer to the black hole ($r_{flare}=6.5m$) 
as shown in Fig. 7 and Fig. 11. 
Fig. 11 shows three distinct maxima with the middle maximum due
to light focusing being dominant.
As in the previous cases, such an effect could be observable by 
future high throughput spectrometers.
The magnitude of the energy shift of the main maximum is related to the 
inclination of the accretion disc (e.g. compare Fig. 3 and Fig. 4) and 
also to the distance of the flare from the centre.
The latter effect is demonstrated for example in Fig.3, Fig. 7 and Fig. 8
where the bright $\Lambda$-shaped feature extends gradually to lower
energies as the distance $r_{flare}$ of the flare from the centre decreases.
The same figures also illustrate how the timescale of the sinusoidal
variation changes as a function of $r_{flare}$
with shorter time scales corresponding to faster orbital motion of the
closer flares.
Note how the structures on the reverberation map change for a very close flare
($r_{flare}=3m$; see Fig. 8).
There is no triple maximum on the left hand side of the
$\Lambda$-shaped feature, but the maximum due to focusing emerges on the
opposite side. This is also shown on Fig. 12 where three maxima are
visible but, unlike in the case of more distant flares, in this case the
lowermost and middle maxima are due to gravitational focusing. The lack 
of the uppermost and middle maxima on the left hand side of this feature
is related to the very large overall
gravitational and transverse Doppler redshift. It is also caused
by the faster motion of the flare (and thus the main flux maximum) 
combined with the relatively slow propagation of signals
slowed by the Shapiro delay in the vicinity of the black hole.
In other words, the ratio between
the time it takes for the flare to complete a half of 
one full revolution around the black hole to the signal crossing time
decreases with decreasing
distance of the flare from the centre. The above effects contribute to the 
swamping of these two maxima by the main bright maximum.\\
\indent
We have demonstrated that the very complex behaviour of the lines
is predicted by a relatively simple model of the orbiting flare.
Our results can be applied
to future observations made with XMM and particularly Constellation-X
and may help to understand various processes operating 
in the vicinity of black holes.
These results can be used to put constraints on their masses
and spin parameters, the geometry of the emitting region and
test general relativity in the strong gravity regime. Particularly
interesting features seen on our diagrams are the triple maxima in the 
temporary iron line profiles. The middle maximum is a consequence
of strong light bending by the extreme gravitational fields very close to
the central black hole and is thus a prediction of general relativity.
This may indicate that a complicated temporal 
behaviour of the iron line 
profiles seen in future data may not necessary imply a complicated model 
but could be explained for example in the framework of the orbiting flare
model.

\section{acknowledgments}
MR acknowledges support from an External Research Studentship of Trinity
College, Cambridge; an ORS award; and the Stefan Batory Foundation. MR thanks 
Andrew Fabian and Andrew Young for discussions.

\appendix
\section{the kerr metric}
Space-time in the vicinity of a rotating black hole is described by the 
Kerr metric. In Boyer-Lindquist coordinates this metric reads:
\begin{eqnarray*}
ds^{2} & = & -\left(1-\frac{2r}{\Sigma}\right)dt^{2}-\frac{4ar}{\Sigma}\sin^{2}\theta
dtd\phi+\frac{\Sigma}{\Delta}dr^{2}+\Sigma d\theta^{2}\\
       &   & +\frac{A}{\Sigma}\sin^{2}\theta d\phi^{2} ,
\end{eqnarray*}
where
\[\Sigma=r^{2}+a^{2}\cos^{2}\theta\]
\[\Delta=r^{2}+a^{2}-2r\]
\[A=(r^{2}+a^{2})^{2}-a^{2}\Delta\sin^{2}\theta\]

\section{the constants of motion for the off-axis source on a spatially 
circular orbit}
The photon trajectory may be specified by two constants of motion (the component of angular momentum parallel to the symmetry axis $l$ 
and the Carter constant $Q$) which can be expressed in terms of the direction cosines $e_{\hat{i}}$
of the
 photon momentum ${\bf k}$ and the comoving tetrad $\lambda_{\hat{i}}$.
\[
e_{\hat{i}}=\frac{{\bf k}\cdot\lambda_{\hat{i}}}{({\bf k}\cdot{\bf h}\cdot{\bf k})^{1/2}}
\hspace*{2cm}\hat{i}=r,\theta
\] 
\noindent
where ${\bf g}$ is the Kerr metric tensor and ${\bf h}$ is the transverse 
projecting operator defined by ${\bf h}={\bf g}+{\bf u\cdot u}$ and
${\bf u}$ denotes the four velocity of the source given by:\\
\[
{\bf u}=C(\partial_{t}+\Omega\partial_{\phi})\hspace*{1.5cm}
{\bf u}\cdot {\bf u}=-1
\] 
\noindent
where
\[
C=\left[1-\frac{2r}{\Sigma}(1-a\Omega\sin^{2}\theta)^{2}-(r^{2}+a^{2})^{2}\Omega^{2}\sin^{2}\theta\right]^{-1/2}
\]
and $\Omega$ is the angular velocity of the flare {\sl defined} by:
\[
\Omega\equiv\frac{1}{a+(r\sin\theta)^{3/2}}
\]
Note that such a definition does not violate (for the parameters considered
in Section 3) the obvious condition that the source must follow a timelike
worldline:
\[({\bf u}\cdot {\bf u}=-1) \Rightarrow [g_{tt}+2\Omega g_{\phi t}+
\Omega^{2}g_{\phi\phi}] < 0\]
The necessary components of the comoving tetrad and photon momentum are:
\[
\lambda_{\hat{r}}=\left(\frac{\Delta}{\Sigma}\right)^{1/2}\partial_{r}
\]
\[
\lambda_{\hat{\theta}}=\Sigma^{-1/2}\partial_{\theta}
\]
and
\[
k^{t}=\frac{A}{\Delta\Sigma}(1-l\omega)
\]
\[
k^{r}=\pm\frac{1}{\Sigma}\left[(2r-al)^{2}+\Delta(r(r+2)-L)\right]^{1/2}
\]
\[
k^{\theta}=\pm\frac{1}{\Sigma}\left(L-\frac{l^{2}}{\sin^{2}\theta}+a^{2}\cos^{2}\theta\right)^{1/2}
\]
\[
k^{\phi}=\frac{A}{\Delta\Sigma}\left[\left(\frac{\Sigma-2r}{A}\right)\frac{l}{\sin^{2}\theta}+\omega\right]^{1/2} ,
\]
where $\omega=2ar/A$ is the angular velocity of the frame dragging and 
$L=Q+l^{2}$.
The directional cosines can be expressed as functions of polar $\Psi$ 
and azimuthal $\Phi$ angles in the rest frame of the source 
in the usual way:
\[e_{\hat{r}}=\cos\Psi\]
\[e_{\hat{\theta}}=\sin\Psi\cos\Phi\]
Here we chose the local z axis to point in the $\partial_{r}$ direction
and the local x axis in $\partial_{\theta}$ direction.
We can now combine the above equations to obtain the desired set of two 
equations for the two constants of motion $l$ and $Q$:

\begin{equation}
\Sigma(e_{\hat{r}}^{2}+e_{\hat{\theta}}^{2}){\bf k\cdot h\cdot k}
+\frac{l^{2}}{\sin^{2}\theta}
=\frac{(2r-al)^{2}}{\Delta}+r(r+2)
\end{equation}
\begin{equation}
\frac{e_{\hat{r}}^{2}}{e_{\hat{\theta}}^{2}}(a^{2}\cos^{2}\theta-\frac{l^{2}}{\sin^{2}\theta}+L)+L=\frac{(2r-al)^{2}}{\Delta}+r(r+2)
\end{equation}
All components containing $L$ in Eq. B1 cancel out miraculously and
we obtain relatively simple quadratic equation for $l$ which has the solution:
\[l=-\frac{Y}{2X}+\frac{1}{2}\mbox{sgn}(\pi-\Phi)\left[\left(\frac{Y}{X}\right)^{2}-
4\left(\frac{Z}{X}\right)\right]^{1/2}\]
with
\[
X=\left(\frac{A}{\Delta\Sigma}\right)^{2}
\left[qp(\omega)\left(\frac{A}{\Sigma}\sin^{2}\theta\right)^{-1}+
C^{2}\left[(\Omega-\omega)
q+p(\Omega)\omega\right]^{2}\right]
\]
\[
\hspace*{0.5cm}-\frac{1}{\Sigma}\left(\frac{a^{2}}{\Delta}-
\frac{1}{\sin^{2}\theta}\right)f(\Psi,\Phi)
\]
\[Y=2\left(\frac{A}{\Delta\Sigma}\right)^{2}\left[\omega p(\omega)+C^{2}\left(
\right. \right. \frac{A}{\Sigma}\omega(\Omega-\omega)^{2}q\sin^{2}\theta-p^{2}(\Omega)\omega\]
\[\hspace*{0.5cm} \left.\left. -(\Omega-\omega)p(\Omega)p(-\omega)\right)\right]
+\frac{4ar}{\Delta\Sigma}f(\Psi,\Phi)\]
\[
Z=-\frac{1}{\Sigma}\left(\frac{4r^{2}}{\Delta}+r(r+2)+a^{2}\cos^{2}\theta\right)
f(\Psi,\Phi)\]
\[
\hspace*{0.5cm}+\left(\frac{A}{\Delta\Sigma}\right)^{2}p(\omega)(C^{2}p(\omega)-1) ,
\]
where
\[p(x)=1-\frac{2r}{\Sigma}+\frac{A}{\Sigma}\omega x\sin^{2}\theta,
\hspace*{0.5cm} f(\Psi,\Phi)=\frac{\sin^{2}\Psi\sin^{2}\Phi}{1-\sin^{2}\Psi\sin^{2}\Phi}\]
and $q=1-(2r/\Sigma)$.
\noindent
The second constant of motion $Q=L-l^{2}$ can be easily calculated 
from Eq. B2.

\section{Illumination of the disc and light curve formulae}
The observed flux consists of two elements: the direct flux from 
the flare and the reflected flux from the accretion disc. 
Let us first consider the observed flux $F(E_{d})$ 
as seen by the disc element moving with the accretion flow.
\begin{equation}
F(E_{d})=\frac{E_{d}dN_{E_{d}}}{dS},
\end{equation}
where $E_{d}$ is the photon energy in the matter rest frame and 
$dN_{E_{d}}$ is the number
of photons per erg per second
 which illuminate an element
of surface $dS$ on the accretion disc. 
 Because the number of photons has to 
be conserved and $dt_{d}dE_{d}=dt_{e}dE_{e}$ we have $dN_{E_{d}}=dN_{E_{e}}$,
where the subscript $e$ and $d$ denote the rest frame of the flare
and the disc frame respectively.
 Thus Eq. C1 takes the form:
\[
F(E_{d})=\frac{g_{sd}L_{E_{e}}}{dS}\frac{d\Omega_{e}}{4\pi},
\]
where $d\Omega_{e}=4\pi(dN_{E_{e}}/N_{tot})$ is the solid angle 
in the rest frame of the emitter
which 
corresponds to the surface element $dS$ on the disc,
 $dN_{Ee}$ is the number of photons 
emitted into this solid angle, $g_{sd}$ is the redshift factor and
$L_{E_{e}}$ is the luminosity of the source per unit energy.
Note that in special relativity the transformation law for the solid 
angle is $d\Omega_{e}=g_{sd}^{2}d\Omega_{obs}$. This means that
in this regime we can recover the usual result $F_{tot}\propto g_{sd}^{4}$ 
by integrating the formula for the flux received by the disc element.
Since the surface density of photons
is locally constant we have
$dS=(dN_{E_{e}}/d\tilde{N}_{E_{e}})d\tilde{S}$, where 
$d\tilde{S}=\gamma (A/\Delta)^{1/2}d\phi dr$
is the surface area of the fixed grid element on the disc
and $\gamma$ is the Lorentz factor of the relative motion of the disc element
and the locally non-rotating observer.
For $r\leq r_{ms}$, where $r_{ms}$ is the radius of marginal stability, this
factor is given by:
\[\gamma=(1-V^{(\phi) 2})^{-1/2},\]
where
\[V^{(\phi)}=[A/(\Sigma\Delta^{1/2})](\Omega_{e}-\omega),\hspace*{1.5cm}
\Omega_{e}=1/(r^{3/2}+a),\]
whereas for $r_{h}<r<r_{ms}$, where $r_{h}$ is the radius of the event horizon:
\[\gamma=(1-\tilde{V}^{(r) 2}-\tilde{V}^{(\phi) 2})^{-1/2},\]
where
\[\tilde{V}^{(r)}=\frac{A^{1/2}}{\Delta}\frac{u^{r}}{u^{t}}\]
\[\tilde{V}^{(\phi)}=\frac{A}{\Delta^{1/2}\Sigma}\left(\frac{u^{\phi}}{u^{t}}-
\omega\right).\]
The variables $u^{t}$, $u^{r}$ and $u^{\phi}$ are the components of the 
4-velocity
below the innermost stable orbit (see e.g. Cunningham 1975) and $V^{\phi}$,
$\tilde{V}^{r}$ and $\tilde{V}^{\phi}$ are the components of 3-velocity
relative to the locally non-rotating frame which can be calculated from
$\tilde{V}^{(j)}=(u^{\mu}e_{\mu}^{(j)})/(u^{\nu}e_{\nu}^{(t)})$, 
where $j=r,\phi$
and $e_{\mu}^{(\nu)}$ are the components of the basis of one-forms corresponding
to the set of basis vectors of the locally non-rotating observer 
(see Bardeen, Press, Teukolsky, 1972).
Thus for the luminosity in the form of the power law 
$L_{E_{e}}\propto E_{e}^{-\alpha}$ we get:
\[F(E_{d})\propto S_{E_{d}}=g_{sd}^{1+\alpha}E_{d}
^{-\alpha}\frac{f_{sd}}{\gamma d\phi dr}\left(\frac{\Delta}{A}\right)^{1/2} ,
\]
where $f_{sd}=d\tilde{N}_{E_{d}}/N_{tot}$ is the fraction of $N_{tot}$
which corresponds to the grid element 
defined by $d\phi$ and $dr$. Assuming isotropic reflection,
the final expressions for the reflected component of the flux read:
\[F_{R}(t)=\sum_{n}F_{R}^{(n)}(t),\]
\noindent
where $F^{(n)}_{R}(t)$ is the contribution to the flux from the flare
at n-th position given by:
\[
F^{(n)}_{R}(\tilde{t})=\frac{1}{\Delta \tilde{t}}
 \sum_{\tilde{t}\rightarrow \tilde{t}+\Delta \tilde{t}}C(E_{o1},E_{o2},\alpha)
\int \int_{E_{o1}}^{E_{o2}}g_{do}^{3}I_{R}(E_{o}/g_{do})
dE_{o}\frac{dxdy}{r_{o}^{2}},
\]
where $\tilde{t}$ is the sum of the time $\tilde{t}_{sd}$ 
it takes for the photon 
emitted at the source to reach a given element on the disc and the 
photon arrival time $\tilde{t}_{do}$ from that particular point on the disc to 
the observer. $g_{do}$ is the redshift factor for the disc-to-observer case, 
$C$ is the proportionality coefficient which depends on the energy index 
and the observed spectral band, $\Delta \tilde{t}$ is the width of 
the time bin and
\[I_{R}(E_{o}/g_{do})=\frac{1}{\pi}a_{R}(\xi,E_{d})F(E_{d}),\]
where $a_{R}(\xi,E)$ is the albedo of the reflecting material.
We used the transformation law for the intensity 
$I_{E_{o}}=g_{sd}^{3}I_{E_{e}}$ in the expression for $F^{(n)}_{R}$ above.
Note that in the formula for $F^{(n)}_{R}$ we include contributions from all
photon isodelay regions on the disc. We then interpolate $F^{(n)}_{R}$ 
at a given time $t$ for all flare positions and sum these contributions
to get the reflected flux $F_{R}(t)$. We use a very similar method to
calculate the iron line spectra (appropriate formulae are given in the 
main text).\\
\indent
Analogously, we obtain the formula for the
 direct component of the flux in the following form:
\[
F_{D}=C(E_{o1},E_{o2},\alpha)g_{so}^{1+\alpha}\frac{f_{so}}{\Delta x\Delta y}\int_{E_{o1}}^{E_{o2}}
E_{o}^{-\alpha}dE_{o} ,
\]
where $f_{so}$ is the fraction of the total number of emitted photons 
intersecting the observer's image plane which has 
the surface area $\Delta x\Delta y$. Note that the formulae 
for $F_{R}$ and $F_{D}$ have
the same proportionality coefficient which enables us to compare the relative
flux contributions.
\end{document}